\documentclass[sort&compress
    ,final            
  ]
  {aipproc}

\layoutstyle{6x9}

\begin{document}

\title{$J/\psi$ Production in $p+p$, $d+Au$, and $Cu+Cu$ Collisions at RHIC}


\author{V. Cianciolo, for the PHENIX Collaboration}{
  address={Physics Division, Oak Ridge National Laboratory, Oak Ridge,
  Tennessee 37831, USA, \newline email: {\tt cianciolotv@ornl.gov}}
}

\begin{abstract}
PHENIX results for $J/\psi$ production in $p+p$, $d+Au$, and $Cu+Cu$
collisions at $\sqrt{s_{NN}}=200$\,GeV are presented. 
\end{abstract}

\maketitle


\section{Introduction}

$J/\psi$'s are an interesting probe of the extremely hot and dense
matter produced in heavy ion collisions at the Relativistic Heavy Ion
Collider (RHIC). Suppression of $J/\psi$ production in heavy ion
collisions (relative to production in $p+p$ collisions scaled by the
number of binary nucleon-nucleon collisions ($N_{coll}$), which is
expected for point-like processes), was predicted by Matsui and
Satz~\cite{Matsui:1986} to be an unambiguous signature for the
formation of the quark-gluon plasma due to Debye screening by free
color charges. Such a suppression has indeed been observed by the NA50
experiment at CERN~\cite{NA50a,NA50b}. However, interpretation has
been made difficult by subsequent work demonstrating numerous
mechanisms that can suppress $J/\psi$ formation~\cite{vogt2000},
including initial parton energy loss, shadowing, cold nuclear matter
absorption, and co-mover absorption. In addition to these effects, the
produced charm quark density at RHIC may allow for $J/\psi$ production
via recombination~\cite[and references therein]{recombination}, thus
countering suppression of prompt production. The mechanism leading to
the large final-state charm-quark energy loss observed at
RHIC~\cite{ppg056} may also affect $J/\psi$ production.

Unraveling this rich structure will require measurement of $J/\psi$
production as a function of numerous independent variables. In this
paper we present current PHENIX results for $J/\psi$ production at
$\sqrt{s_{NN}}=200$\,GeV versus transverse momentum ($p_T$), rapidity
($y$), and system size for light systems ($p+p$, $d+Au$ and $Cu+Cu$).

\section{Results and Summary}

PHENIX~\cite[and references therein]{PHENIXnim} has measured $J/\psi$
production in $p+p$~\cite{ppg017,ppg038}, $d+Au$~\cite{ppg038},
$Cu+Cu$~\cite{PereiraQM05}, and
$Au+Au$~\cite{ppg018,PereiraQM05,TakuPANIC} collisions at
$\sqrt{s_{NN}} = 200$\,GeV. Measurements at forward and
backward rapidity ($1.2 < |y| < 2.2$) are made with the Muon Arm
spectrometers ($J/\psi\rightarrow\mu^+\mu^-$), measurements at
mid-rapidity ($|y|<0.35$) are made with the Central Arm spectrometers
($J/\psi\rightarrow e^+e^-$).

The left panel of Figure 1 shows rapidity distributions for $J/\psi$
production in $p+p$ collisions. This measurement, which goes down to
$p_T = 0$, is consistent with predictions of the color octet model
(COM)~\cite{COM}, as observed by CDF~\cite{CDFJpsi}. Recall that
$J/\psi$ polarization at high $p_T$, predicted by the COM, has not
been observed~\cite{JpsiPol}. The right panel of Figure 1 shows the
nuclear modification factor ($R_{AB} =
\frac{dN^{AB}/dy}{N_{coll}^{AB}dN^{pp}/dy}$) in $d+Au$ collisions,
which is expected to be 1 in the absence of nuclear effects. Only cold
nuclear matter effects are expected in $d+A$ collisions, and these are
seen to be modest. Although these results are consistent with
predictions incorporating shadowing and nuclear absorption they do not
scale with $x_2$ (the momentum fraction in the gold nucleus) when
comparing with lower energies, as would be expected for
shadowing~\cite[and references therein]{ppg038}. The left panel of
Figure 2 shows the nuclear modification factor in $Cu+Cu$ collisions
vs.~collision centrality (quantified by $N_{part}$, the number of
participants as determined with a Glauber
calculation~\cite{glauber}). The suppression increases smoothly to a
factor of two for the most central collisions. This suppression is
consistent with the suppression observed by NA50 and is reasonably
well-reproduced by models incorporating only cold-nuclear matter
effects and by models which incorporate nearly complete suppression of
initial-state $J/\psi$'s plus subsequent formation via recombination
mechanisms~\cite[and references therein]{PereiraQM05}. Models which
successfully reproduced NA50 results overpredict the suppression at
RHIC if recombination is not invoked. The right panel of Figure 2
shows the $J/\psi$ rapidity distributions for $p+p$ collisions and
$Cu+Cu$ collisions of different centrality. These distributions do not
appear to change shape. In particular they do not get significantly
narrower as is predicted for current $J/\psi$ recombination
models~\cite{recombination}. Satz~\cite{satz_quarko} has recently
suggested that the similarity in suppression observed by PHENIX and
NA50 is caused by nearly complete melting of $J/\psi$ precursors
($\psi\prime$ and $\chi_c$) while directly produced $J/\psi$'s remain
intact. No recombination is invoked in this model.

$J/\psi$'s hold promise as sensitive probes of hot nuclear matter, but
the probe requires better calibration. In particular, better
measurements in $p+p$ collisions, including polarization measurements,
are needed to understand the initial production mechanism; better
measurements in $d+Au$ collisions are necessary to understand effects
of cold nuclear matter; better measurements of rapidity, $p_T$, and
reaction plane are needed to distinguish between competing final state
effects; measurements of other quarkonia states, particularly
$\psi\prime$ and $\chi_c$ are necessary to understand feeddown
contributions.


\begin{figure}
  \includegraphics[height=.3\textheight]{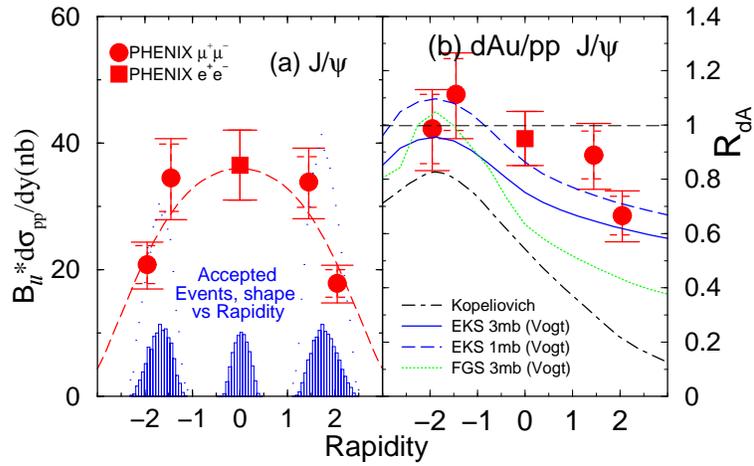}
  \caption{Left: Differential $J/\psi$ cross section (times di-lepton branching
  ratio) vs.~$y$ in $p+p$ collisions. Right: Nuclear modification ratio
  for $J/\psi$ production in $d+Au$ collisions~\cite[and references therein]{ppg038}.}
\end{figure}

\begin{figure}
  \includegraphics[height=.25\textheight]{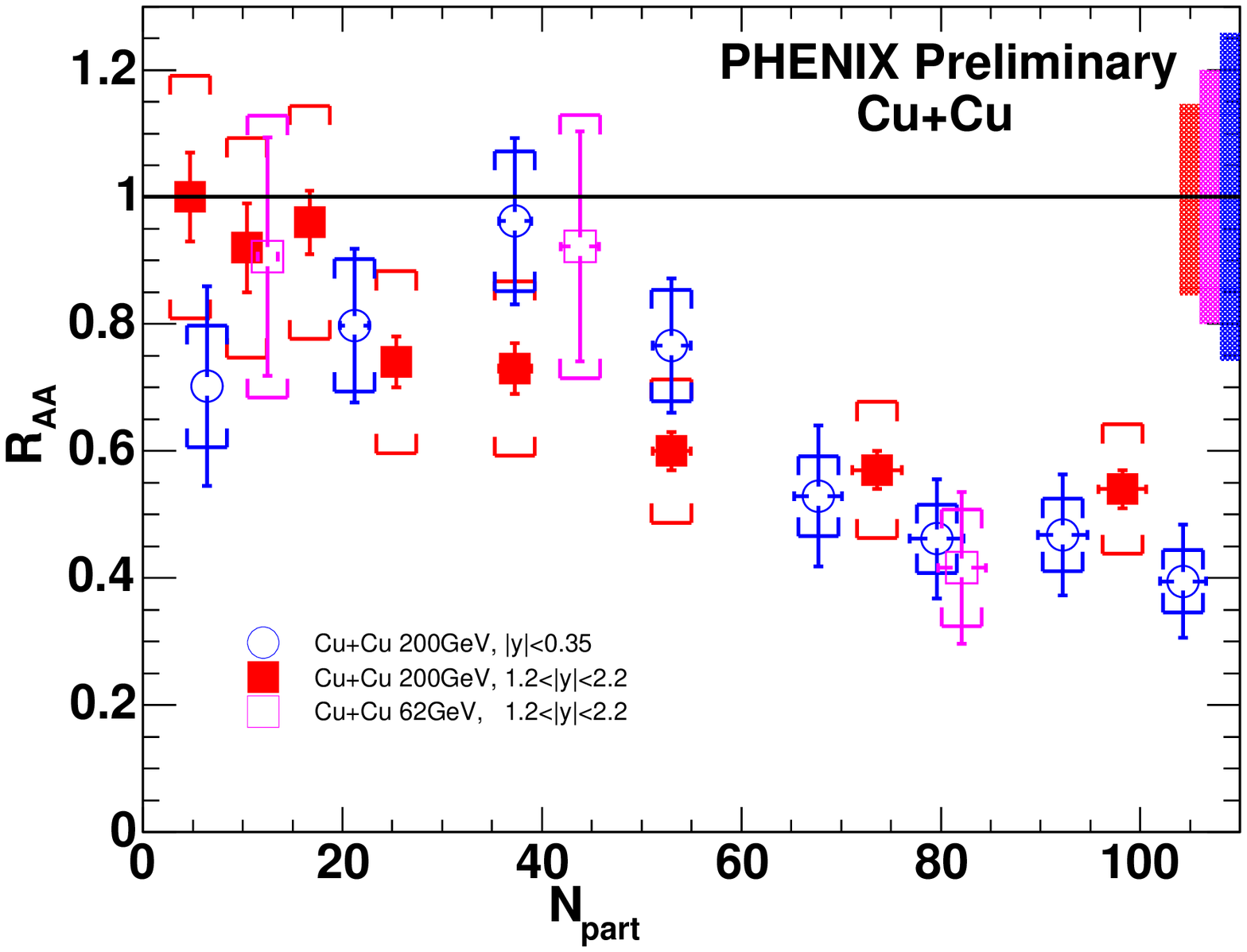}
  \includegraphics[height=.25\textheight]{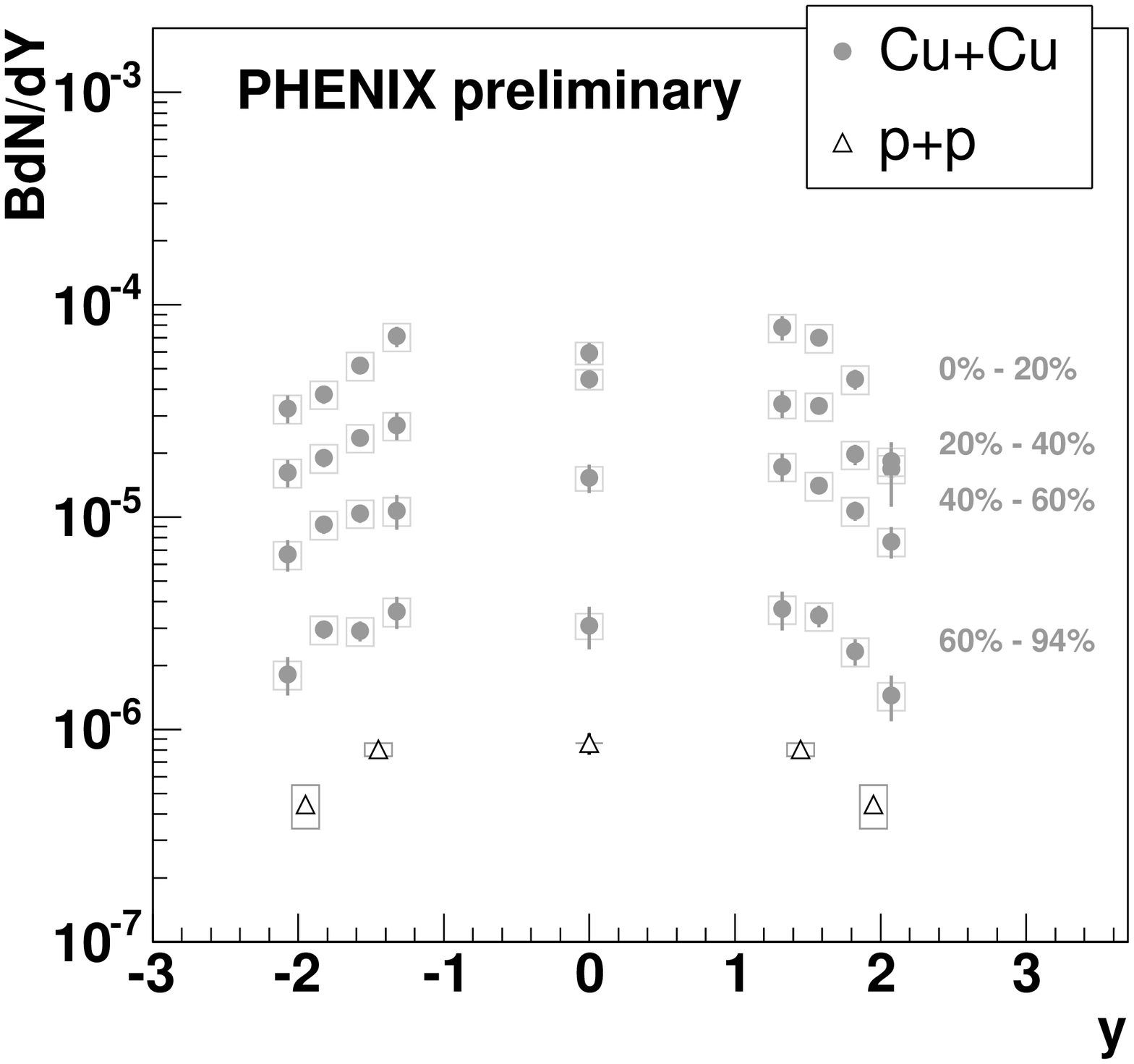}
  \caption{Results for $Cu+Cu$ collisions. Left: Nuclear modification
  factor. Right: Rapidity distributions for indicated
  centrality selections. Results for $p+p$ collisions are shown for comparison.}
\end{figure}





\begin{theacknowledgments}
This work was sponsored by the Division of Nuclear Physics,
U.S. Department of Energy, under contract DE-AC05-00OR22725 with
UT-Battelle, LLC (Oak Ridge National Laboratory).
\end{theacknowledgments}



\bibliographystyle{aipproc}   

\bibliography{mypanic}


\end{document}